\documentclass[12pt]{article}
\usepackage{amsfonts,amssymb}

\date{January 15, 2001}

\def\be{\begin{equation}}
\def\ee{\end{equation}}
\def\bear{\begin{eqnarray}}
\def\eear{\end{eqnarray}}
\def\nn{\nonumber}

\def\dl{\eta}

\newcommand{\px}[1]{{\partial_{#1}}}

\newcommand{\tr}[1]{{\mbox{tr}\{{#1}\}}}          
\newcommand{\com}[2]{{\lbrack {#1},{#2}\rbrack}}  
\newcommand{\bra}[1]{{\langle {#1}|}}
\newcommand{\ket}[1]{{|{#1}\rangle}}

\newcommand{\MR}[1]{{\mathbb{R}^{#1}}}               
\newcommand{\MC}[1]{{\mathbb{C}^{#1}}}               

\def\a{{\alpha}}
\def\b{{\beta}}

\def\u{{\mu}}

\def\z{{\zeta}}
\def\th{{\theta}}

\def\lam{{\lambda}}

\def\bz{{\overline{z}}}
\def\pxz{{\partial}}
\def\bpx{{\overline{\partial}}}

\def\Hil{{\cal H}}             
\def\hH{{\hat{H}}}             

\def\hPhi{{\hat{\Phi}}}
\def\hPsi{{\hat{\Psi}}}
\def\hLam{{\hat{\Lambda}}}
\newcommand{\ha}[1]{{\hat{a}_{#1}}}                 
\newcommand{\hadg}[1]{{{\hat{a}^\dagger_{#1}}}}     
\newcommand{\hadgp}[2]{{{\hat{a}^{\dagger {#2}}_{#1}}}} 
\def\barb{{\overline{\beta}}}
\def\bzeta{{\overline{\zeta}}}
\def\hU{{\hat{U}}}
\def\hO{{\hat{O}}}
\def\hP{{\hat{P}}}
\def\hf{{\hat{f}}}
\def\hp{{\hat{p}}}
\def\hx{{\hat{x}}}
\def\hy{{\hat{y}}}

\begin{document}

\begin{titlepage}
\titlepage
\rightline{hep-th/0101095, PUPT-1971}
\rightline{\today}
\vskip 1cm
\centerline{{\Huge A Note on Intersecting and Fluctuating Solitons}}
\centerline{{\Huge in 4D Noncommutative Field Theory}}
\vskip 1cm
\centerline{Aaron Bergman\footnote{\tt abergman@princeton.edu},
Ori J. Ganor\footnote{\tt origa@viper.princeton.edu} and
Joanna L. Karczmarek\footnote{\tt joannak@princeton.edu}}
\vskip 0.5cm

\begin{center}
Department of Physics, Jadwin Hall \\
Princeton University \\
NJ 08544, USA
\end{center}

\abstract{
We examine the intersections, fluctuations and deformations of
codimension two solitons in field theory on
noncommutative $\MR{4}$, in the limit of large noncommutativity.
We find that holomorphic deformations are zero modes of flat branes, 
and we show
that there is a zero mode localized at the intersection of two solitons.
}
\end{titlepage}



\section{Introduction}\label{intro}
Field theories on noncommutative spaces possess a surprisingly
rich dynamical structure (see for instance
\cite{CDS}-\cite{MinSei}.)
Recently, soliton solutions of scalar field theory on noncommutative
$\MR{n}$ ($n$D NCFT)
have been found
in the limit of large noncommutativity \cite{GoMiSt}.
The results of \cite{GoMiSt} have been extended to gauge theories
in
\cite{GroNekII}-\cite{Pilo:2000nc}.

These results are particularly exciting in light of the application
to D-branes in string theory
\cite{HK}-\cite{HM}.

In this paper we will study the fluctuations of a two-dimensional
soliton (2-brane) in scalar 4D noncommutative field theory
and begin the study of intersecting 2-branes in this theory.
On the geometrical level, a plane in four dimensions can be deformed
into a curved minimal area surface, and
two intersecting planes can be deformed into a single smooth minimal
area surface. In string theory, the latter phenomenon is related to
a zero-mode which appears at the intersection of two D-branes \cite{SenU}.

The purpose of these notes is to:
\begin{enumerate}
\item
Study the small deformations of planar 2-branes.
\item
Identify the classical solutions corresponding to intersecting
2-branes.
\item
Identify the zero-mode corresponding to
a deformation into a smooth surface.
\end{enumerate}

In principle, the zero mode at the intersection
of two 2-branes might not correspond to an exact flat
direction because of the existence of a quartic (or higher) potential.
We will not explore this issue directly but we will study deviations
from the linear equations of motions in the case of small fluctuations
of a flat 2-brane.

Following \cite{GoMiSt}, we will work in the large noncommutativity
limit but include the kinetic energy to first order.


The paper is organized as follows.
In section (\ref{geom}) we review the geometry of the deformation
of intersecting planes. In section (\ref{onebr}) we review the
constructions of \cite{GoMiSt} and study the deformation modes
of a single 2-brane in 4D scalar NCFT.
We will show that half of the deformation modes correspond to deformations
of the flat 2-brane into a holomorphic curve embedded in $\MR{4}$.
The other half correspond to anti-holomorphic fluctuations.
In section (\ref{FlucFlat}) we will study the holomorphic and anti-holomorphic
fluctuations to higher order. In particular cases, we obtain deviations from
pure holomorphicity at $7^{th}$ order!
In section (\ref{twobr}) we describe the solution corresponding to 
two intersecting
branes and study the zero-modes that correspond to their deformations. In
section (\ref{disc}) some extensions to the case of multiple branes and
more dimensions are discussed. We also briefly comment on the situation with
$U(\infty)$ gauge fields.


\section{Classical Geometry}
\label{geom}
We will consider surfaces in $\MR{4}$ that can be described
by a holomorphic equation when $\MR{4}$ is identified with $\MC{2}$.
Such surfaces have a minimal area in the sense that small deformations
of the surface, keeping the boundary conditions at infinity intact,
never decrease the area.
Let the coordinates be:
$$
z_k \equiv x_k + i y_k,\qquad k=1,2.
$$

Consider first a surface that spans the $z_2$-direction and is
given by the equation $z_1 = 0$.
Small holomorphic
deformations are described by $z_1 = \epsilon f(z_2)$
with $f(z) = \sum_{n=0}^\infty c_n z^n$ a holomorphic function.

Now consider adding a second surface spanning the $z_1$ direction,
with the equation $z_2 = 0$. The two surfaces can be represented together
by the equation $z_1 z_2 = 0$.
This reducible surface can
be deformed into a smooth irreducible surface given by $z_1 z_2 = \zeta$
where $\zeta$ is a complex number.
This is the only holomorphic deformation
of the singular surface $z_1 z_2 = 0$ that preserves the boundary
conditions $z_1\rightarrow 0$ as $|z_2|\rightarrow\infty$ and
$z_2\rightarrow 0$ as $|z_1|\rightarrow\infty$.

In this case, we see that the possible deformations are given
by $z_1 = \epsilon f(z_2)$ where $f(z) = \sum_{n=-1}^\infty c_n z^n$
is allowed to have a simple pole at $z=0$.
More generally, if we add $r$ surfaces given by the planes
$z_2 = \xi_j$ ($j=1\dots r$), we can have deformations
$z_1 = \epsilon f(z_2)$ where $f$ is a meromorphic function
that is allowed to have simple poles at $\xi_1,\dots,\xi_r$.
If we add a surface $z_2=0$ with multiplicity $k$, then $f(z)$
is allowed to have a pole of $k^{th}$ order at the origin.


\section{A Single Brane and its Fluctuations}
\label{onebr}
In this section we will construct solitons of noncommutative
scalar field  theory along the lines of \cite{GoMiSt}.

\subsection{The Soliton}
Let us review the construction of \cite{GoMiSt} for a single
codimension-2 brane in the theory with action:
$$
\int \lbrack (\px{\u}\Phi)^2 + V(\Phi)\rbrack.
$$
Here:
$$
V(\lam) = \sum_{n=2}^\infty a_n \lam^n,\qquad
V(\Phi) = a_2 \Phi\star\Phi + a_3 \Phi\star\Phi\star\Phi + \cdots
$$
We take spacetime to be commutative and define the $\star$-product as:
$$
\Phi\star\Psi\equiv
\Phi
      e^{ \frac{i\th}{2}\frac{\stackrel{\leftarrow}{\partial}}{\partial x_1}
            \frac{\stackrel{\rightarrow}{\partial}}{\partial y_1}
         -\frac{i\th}{2}\frac{\stackrel{\leftarrow}{\partial}}{\partial y_1}
            \frac{\stackrel{\rightarrow}{\partial}}{\partial x_1}}
    \Psi
$$
So that:
$$
x_1\star y_1-y_1\star x_1 = i\theta.
$$
We take the limit $\th\rightarrow\infty$.
After a rescaling of the coordinates, the kinetic term
is of order $1/\th$ and can be neglected.  For now, the $x_2,y_2$
coordinates are still commutative.

We set $z_1 = x_1 + i y_1$ and
define a Hilbert
space $\Hil_1$ with the harmonic oscillator basis,
     $\ket{n}$ for $n=0,1,\dots$,
such that $\hadg{1}\ket{n} = \sqrt{n+1}\ket{n+1}$
and $\ha{1}\ket{n} = \sqrt{n}\ket{n-1}$.
If $\Phi$ is a function of $x_1$ and $y_1$, the Weyl formula
transforms it into an  operator on this Hilbert space:
$$
\hPhi\equiv \frac{1}{2\pi}\int d^2\zeta\,\Phi(z_1,\bz_1)
     e^{i\zeta\bz_1-i\bzeta z_1}.
$$
Then $z_1\rightarrow \sqrt{2\th}\ha{1}$ and 
$\bz_1\rightarrow\sqrt{2\th}\hadg{1}$.
{}From now on, $\Phi,\Psi,\dots$ will denote ordinary functions
and $\hPhi,\hPsi,\dots$ will denote the corresponding operators.

Let us assume that $V(\Phi)$ has a minimum at $\lam\neq 0$.
One can then construct a soliton by setting:
$$
\hPhi = \lam \hP,\qquad \hP^2 = \hP.
$$
The operator $\hPhi$ satisfies $V(\hPhi) = V(\lam) \hP$ and
hence $V'(\hPhi)=0$.
The corresponding (Weyl transformed) solution, $\Phi$, is constant in the
$z_2$ direction. For any unitary
operator, $\hU$, V'($\hU^{\dagger} \hPhi \hU$) is also zero.

If we now include the kinetic term,
only the operators of the form
$$
\hP = \ket{\a}\bra{\a},\qquad
\ket{\a} \equiv e^{\a \hadg{1} - \overline{\a} \ha{1}} \ket{0},
$$
corresponding to projections onto a
coherent state of the harmonic oscillator,
remain as good solitons.
To see this we can write the kinetic energy as
$$
K = -\frac {1}{2\theta^2}
\tr{\com{\hx_1}{\hPhi}^2 + \com{\hy_1}{\hPhi}^2},\qquad
\hx_1\equiv\sqrt{\frac{\th}{2}}(\ha{1}+\hadg{1}),\quad
\hy_1\equiv -i\sqrt{\frac{\th}{2}}(\ha{1}-\hadg{1}).
$$
For $\hP = \ket{\phi}\bra{\phi}$, we find
$$
\frac{\theta^2}{\lambda^2} K = \Delta x_1^2 + \Delta y_1^2
$$
where
$$
\Delta x_1^2 = \bra{\phi} \hx_1^2 \ket{\phi} -\bra{\phi} \hx_1\ket{\phi}^2,
\qquad
\Delta y_1^2 = \bra{\phi} \hy_1^2 \ket{\phi} -\bra{\phi} \hy_1\ket{\phi}^2
$$
are the uncertainties in $\hx_1$ and $\hy_1$.
Now we can see that the coherent states, $\ket{\a}$,
minimize the kinetic energy. This is because:
$$
\Delta x_1^2 + \Delta y_1^2 \ge 2 \Delta x_1 \Delta y_1 \ge 1,
$$
and the equalities hold only for a coherent state.
Thus, in the space of all possible unitary transformations,
$\hU$, acting on $\hPhi$, the kinetic energy has flat directions
corresponding to translating the brane rigidly in
the $z_1$ direction.

Now, let us add two extra noncommutative directions:
$$
x_1\star y_1-y_1\star x_1 =x_2\star y_2 -y_2\star x_2 = i\th.
$$

As with $z_{1}$, $z_2$ corresponds to an operator on a
Hilbert space $\Hil_2$.  $\Phi$, as a function of $x_1,~y_1,~
x_2$ and $y_2$, corresponds to an operator on the Hilbert
space $\Hil = \Hil_1 \otimes \Hil_2$.  $\Hil$ has a basis
$\ket{N,n}$ defined by
$\hadg{1}\ket{N,n} = \sqrt{N+1}\ket{N+1,n},$
$\ha{1}\ket{N,n} = \sqrt{N}\ket{N-1,n},$
$\hadg{2}\ket{N,n} = \sqrt{n+1}\ket{N,n+1}$
and $\ha{2}\ket{N,n} = \sqrt{n}\ket{N,n-1}.$ This is just
the tensor product of the harmonic oscillator eigenstates in each
Hilbert space. The soliton described above, corresponding to a codimension-2
brane with $z_1=0$, is now described by $\hPhi = \lambda \hP_1$,
where $\hP_1$ is given by
\be
\hP_1 = \sum_{n=0}^\infty \ket{0,n} \bra{0,n}.
\ee
The codimension-2 brane with $z_2=0$ is similarly given by
$\hPhi = \lambda \hP_2$,
\be
\hP_2 = \sum_{N=0}^\infty \ket{N,0} \bra{N,0}.
\ee

\subsection{Unitary Fluctuations}

We now consider the soliton given by $\hU^{\dagger} \hP_1 \hU$,
where $\hU$ is some unitary operator on $\Hil = \Hil_1 \otimes \Hil_2.$
We are interested in the kinetic energy as a function of $\hU$.
This is more involved than before, so we will work only to second
order with
$$
\hU = e^{i\epsilon\hLam} = 1 + i\epsilon \hLam +
-\frac{1}{2} \epsilon^2 \hLam^2 +
\mathcal{O}(\epsilon^3)
$$
for $\epsilon$ real and small and $\hLam$ Hermitian.
Define
$$
\hLam \ket{0,j} = \sum_{I,i} b^j_{Ii} \ket{I,i}.
$$
Following \cite{GoMiSt},
we now obtain the effective Hamiltonian for small fluctuations
of the brane.
In the operator language, the kinetic energy is:
\bear
K &=& -\frac{1}{2\th^2}\sum_{k=1}^2\tr{
\com{\hx_k}{\hPhi}^2 + \com{\hy_k}{\hPhi}^2}
=
\frac{1}{\th^2}\sum_{k=1}^2\tr{
\com{\ha{k}}{\hPhi} \com{\hPhi}{\hadg{k}}}
\nn\\ &=&
\frac{2}{\th}\tr{
\hPhi\hH\hPhi -\sum_{k=1}^2 (\hPhi \ha{k}\hPhi) (\hPhi \hadg{k}\hPhi)}
\nn
\eear
where $\hH$ is the harmonic oscillator Hamiltonian,
$$
\hH \equiv \sum_{k=1}^2 \left(\hadg{k}\ha{k} + \frac{1}{2}\right).
$$

Any projection operator, $\hat{A}$, such as our soliton,
projects onto a subspace,
$\mathcal{H}_{A}$, of the Hilbert space $\mathcal{H}$. Let $\ket{i}$,
$i \in \mathcal{S}$, be a basis for $\mathcal{H}_{A}$. Then we can
write the kinetic energy as
\be
\label{kehilb}
K = \frac{\lambda^2}{\th^2}\left(\sum_{i\in\mathcal{S};k=1,2}\bra{i}\hadg{k}
\ha{k} + \ha{k} \hadg{k} \ket{i} - 2\sum_{i,j\in\mathcal{S};k=1,2}
|\bra{i} \hadg{k}\ket{j}|^{2}\right)
\ee
\noindent This form is sometimes more useful for calculation.

For fluctuations about $P_1$, to second order in $\epsilon$ we obtain:
\bear
\frac{\theta} {2 \lambda^2} K &=& \sum_k (2k+2) +
2 \epsilon^2 \Bigg[
\sum_{I \ge 2, j,k \ge 0} (I+k-j) |b^k_{Ii}|^2 -
\sum_{j,k \ge 0} |b^k_{1j}|^2 \nn\\
&-& \sum_{j \ge 0} (j+1) \left (
1 - \sum_{I \ge 1, i \ge 0} |b^j_{Ii}|^2 - \sum_{I \ge 1, i \ge 0}
|b^{j+1}_{Ii}|^2
\right ) \nn\\
&-& \sum_{I,i,j \ge 1} \sqrt{i(j+1)}
\left (
\bar b^j_{I,i-1} b^{j+1}_{I,i} + b^j_{I,i-1} \bar b^{j+1}_{I,i}
\right )\Bigg].
\eear

This can be rearranged to the positive definite form:
\bear
\frac{\theta} {2 \lambda^2} K = T + 2\epsilon^2 \Bigg[
\sum_{I \ge 2; i,k \ge 0} I |b^k_{Ii}|^2 +
\sum_{I \ge 1; i \ge 0} i |b^0_{Ii}|^2 \nn\\
+ \sum_{I \ge 1; j,k \ge 0}
\left|\sqrt{k+1} b^k_{Ij} - \sqrt{j+1} b^{k+1}_{I,{j+1}}\right|^2
\Bigg].
\label{originalKE}
\eear
Here, $T$ is an infinite constant corresponding to the
zero-point energy of the infinite brane. The massless
modes must satisfy
\bear
b^m_{Ii} = 0\,({\mbox{for $I\ge 2$}}),\nn\\
b^0_{1,i} = 0\,({\mbox{for $i\ge 1$}}), \\
\sqrt{m+1}b^m_{1,n} = \sqrt{n+1} b^{m+1}_{1,n+1}.
\eear
The solution to these constraints is
$$
b^m_{1,n} = \left\{\begin{array}{ll}
0 & {\mbox{for $m < n$}} \\
\sqrt{\frac{m!}{n!}} c_{m-n} &
     {\mbox{for $m \ge n$}} \\
\end{array}\right.
$$
where $c_m$ ($m=0,1,\dots$) are arbitrary constants. Note that
when looking at the original form of the kinetic energy (\ref{originalKE}), 
we are cancelling two divergent sums. If we demand that all sums converge,
the following solution is not legitimate. Throwing caution to the
wind, we define the entire holomorphic function
$f(\zeta) = \sum_m c_m \zeta^m$.
$\hLam$ can then be written as:
$$
\hLam = \hadg{1} f(\ha{2}) +\ha{1} f(\ha{2})^\dagger
+\mathcal{O}(\epsilon^2),
$$
and the transformed soliton is:
$$
\hPhi = \lambda \hU^\dagger \hP_1 \hU, \qquad \hU =
e^{i\epsilon(\hadg{1} f(\ha{2}) + \ha{1} f(\ha{2})^\dagger)} 
+\mathcal{O}(\epsilon^2).
$$
Physically, this is interpreted as a deformation of
the brane from $z_1=0$ to $z_1= \sqrt{2\th}\epsilon 
f(\frac{z_2}{\sqrt{2\th}})$.
We can now understand the divergences in this solution
as stemming from the fact that a nonconstant entire function
cannot be bounded and, as such, these are infinitely large
deformations of the brane.  If we cut off the sums to force them
to be finite, we can still understand these as
local approximate zero modes.  Another way to understand local
behavior is to begin with the equations of motion which follow from
the above kinetic energy.  This allows us to directly study localized
fluctuations. We will examine this further in section (\ref{FlucFlat}).

We can rearrange the terms in the kinetic energy into the
following (also positive definite) form:
\bear
\frac{\theta} {2 \lambda^2} K = T + 2\epsilon^2 \Bigg[
\sum_{I \ge 2, i,k \ge 0} I |b^k_{Ii}|^2 +
\sum_{I \ge 1, k \ge 0} k |b^k_{I0}|^2 \nn\\
+ \sum_{I \ge 1, j,k \ge 0}
\left|\sqrt{j+1} b^k_{Ij} - \sqrt{k+1} b^{k+1}_{I,{j+1}}\right|^2
\Bigg].
\eear
Repeating the above analysis, we find that the massless
modes for this form of the kinetic energy are
$$
b^m_{1,n} = \left\{\begin{array}{ll}
0 & {\mbox{for $m > n$}} \\
\sqrt{\frac{n!}{m!}} c_{n-m} &
     {\mbox{for $m \le n$}} \\
\end{array}\right..
$$
Taking again $f(\zeta) = \sum_m c_m \zeta^m$, we
obtain
$$
\Lambda = z_1^\dagger f(z^\dagger_2) - z_1 f(z^\dagger_2)^\dagger.
$$
This corresponds to a deformation of the brane from $z_1=0$ to
$z_1=\sqrt{2\th}\epsilon f(\frac{\bz_2}{\sqrt{2\th}})$, an 
antiholomorphic deformation.

\section{Small fluctuations of a flat brane at higher orders}
\label{FlucFlat}

What happens to the zero modes that describe the holomorphic
and anti-holomorphic fluctuations at higher orders?

A ``classical'' 2D (static) membrane in $\MR{4}$ is described
by the equation of motion that states that the area should be minimal under
local deformations. At large distances,
the solitons in noncommutative field theory also look like
2D membranes, and we will assume that the curvature, $R$, of these
solitonic membranes is much smaller than the scale set by
the noncommutativity, $R\ll \th^{-2}$.
In this section we will set $\th=\frac{1}{2}$.

These noncommutative solitons differ from the classical membrane
in two major ways:
\begin{itemize}
\item
The antisymmetric 2-form that determines the noncommutativity
specifies a preferred complex structure. Thus the $SO(4)$ symmetry
of $\MR{4}$ is broken to $U(2)$.
This suggests that deforming a flat soliton by an anti-holomorphic
deformation into a curve of the form $z_1=\epsilon f (\bz_2)$
might not be an exact solution.

\item
The effective action of the soliton might receive curvature
dependent corrections even for a holomorphic deformation $z_1 
=\epsilon f(z_2)$.

\end{itemize}

In this section we will study both these questions.
We set
$$
\hPhi \equiv e^{-i\hLam} \hP_1 e^{i\hLam},\qquad
\hLam \equiv \epsilon \hadg{1} \hat{f}(\ha{2},\hadg{2})
+\epsilon\ha{1}\hat{f}(\ha{2},\hadg{2})^\dagger,
$$
and study the corrections to the equations of motion.
We continue to work in the approximation that $\th$ is large.

After we find the corrections to the operator $\hPhi$ in an 
$\epsilon$ expansion,
we will translate the operator $\hPhi$ into a function 
$\Phi(z_1,z_2,\bz_1,\bz_2)$
via the Weyl transformation:
\bear
\Phi(z_1,z_2,\bz_1,\bz_2) &=&
\frac{1}{\pi^2}\int \prod_{k=1}^2 d^2\zeta_k\,
e^{i\sum_{k=1}^2\zeta_k\bz_k+i\sum_{k=1}^2\overline{\zeta}_k z_k}
\tr{e^{-i\sum_{k=1}^2\zeta_k\hadg{k}-i\sum_{k=1}^2\overline{\zeta}_k\ha{k}}\hPhi}.
\nn
\eear
We will then solve for the maximum of $\Phi$ for a given $z_2$ so as to find
the equation for the curve that is the approximate macroscopic description
of the soliton.
This is an equation of the form $z_1 = \varphi(z_2,\bz_2)$.
To lowest order in $\epsilon$ we always
obtain $\varphi = \frac{1}{\sqrt{2}}\epsilon f + 
\mathcal{O}(\epsilon^2)$, where
$f$ is the Weyl transform of $\hf$.
We will be interested in the higher order corrections.

\subsection{The equations of motion}

We now describe this procedure in greater detail. We begin by
examining the equations of motion.
Instead of writing the equations of motion for $\hLam$, it will be 
more convenient
to write the equations for $\hPhi$ directly.
Starting with
$$
\Phi_0 = 2e^{-|z_1|^2} \Longrightarrow
\hPhi_0 = \sum_{n=0}^\infty \ket{0,n}\bra{0,n},
$$
we take the unitary operator $\hU \equiv e^{i\hLam}$ and
define $\hPhi=\hU^\dagger\hPhi_{0}\hU$.
The equations of motion are obtained by minimizing the kinetic
energy that is proportional to:
$$
\sum_{i=1}^2\tr{\com{\ha{i}}{\hPhi}\com{\hadg{i}}{\hPhi}}
$$
with respect to $\hLam$.
However, it will turn out to be more convenient to write an equation
of motion for $\hPhi$.
We must minimize $K$ under the condition $\hPhi\star\hPhi=\hPhi$, so
we insert a Lagrange multiplier, $\chi$, to enforce the constraint.
This gives:
\be
\label{chieqn}
\Delta\hPhi=\chi\star\hPhi+\hPhi\star\chi-\chi
\ee
where
$$
\Delta\hPhi\equiv\sum_{i=1}^2\com{\ha{i}}{\com{\hadg{i}}{\hPhi}}.
$$
In general, a Hermitian operator, $\hO$, that can be written as
$$
\hO=\chi\star\hPhi+\hPhi\star\chi-\chi
$$
satisfies
$$
\hO\star\hPhi=\hPhi\star\hO.
$$
Alternatively, given an operator, $\hO$, that commutes with $\hPhi$ we can
satisfy (\ref{chieqn}) by choosing
$$
\chi = 2\hPhi\star\hO-\hO.
$$
Thus the equations of motion are equivalent to:
\be\label{eqphph}
\hPhi\star\hPhi=\hPhi,\qquad
\com{\Delta\hPhi}{\hPhi}=0.
\ee

\subsection{Anti-holomorphic fluctuations}
In order to further study anti-holomorphic fluctuations, we take:
$$
\hLam = \rho e^{i\phi} \hadg{1} \hadg{2} + \rho e^{-i\phi}\ha{1} \ha{2},
$$
where $\rho$ and $\phi$ are real.
The Weyl transform of $e^{i\hLam}\hP_1 e^{-i\hLam}$ is
\be\label{solphah}
\Phi = 2 e^{-|z_1\cosh\rho -i \bz_2 e^{i\phi}\sinh\rho|^2}.
\ee
The operator, $e^{i\hLam}$, generates an $SO(4)$ rotation of $\MR{4}$
that is not in
$U(2)\subset SO(4)$. Therefore, the maxima of $\Phi$ correspond to
a plane that is not a holomorphic curve in the preferred complex structure
that is determined by the noncommutativity.
However, it is easy to see that $\hPhi$ is still a solution of the equations
of motion (\ref{eqphph}). This is because the Laplacian operator, $\Delta$,
is $SO(4)$ invariant and not just $U(2)$ invariant.

The kinetic energy density along the
soliton given by (\ref{solphah}) is independent of $\rho$.
However, the ``width'' of the soliton is proportional to $1/\cosh 2\rho$.
So, {\it microscopically}, the solitons that correspond to non holomorphic
curves differ from the holomorphic ones in that they are ``thinner''.
It is amusing to note that for $\rho=\infty$, the curve is 
$z_1=e^{i\phi}\bz_2$,
and the width of the soliton is zero.
However, {\it macroscopically}, all the planar solitons have
the same energy density and the microscopic distinction between
different directions probably disappears
because the $SO(4)$ symmetry is restored.
It would be interesting to confirm this
with scattering calculations.

\subsection{Curvature}
Finally, we would like to study higher order corrections to a
holomorphic deformation. For this, we take:
$$
\hLam_1 = \epsilon (\b \hadg{1}\ha{2}^2 + \barb\ha{1}\hadgp{2}{2})
$$
and define $\hPhi = e^{-i\hLam}\hP_1 e^{i\hLam}$. This corresponds to
placing the brane along the curve $z_{1} = \epsilon\beta\z_{2}^{2}$.
We wish to calculate:
$$
\hat{\Xi}_1\equiv
\com{\Delta\hPhi}{\hPhi}.
$$
For these values, we have:
$$
\hat{\Xi}_1 =-4\epsilon^3 (\b^2\barb\hadg{1}\hP_1 \ha{2}^2
-\b\barb^2\hP_1 \ha{1}\hadgp{2}{2}) + \mathcal{O}(\epsilon^4).
$$
In order to satisfy the equations of motion, this should be zero.
Towards that end, we can cancel the $\epsilon^{3}$ term by augmenting
$\hLam_{1}$ to:
$$
\hLam_2 \equiv
\epsilon (\b\hadg{1}\ha{2}^2 +\barb\ha{1}\hadgp{2}{2})
-\frac{4}{3}\epsilon^3 (\b^2\barb\hadg{1}\hadg{2}\ha{2}^3
+\barb^2\b\ha{1}\hadgp{2}{3}\ha{2}).
$$
This should not be considered a modification of the equation
of motion for the fluctuation $f(\ha{2},\hadg{2})$ because, at the 
current order
of approximation in $\epsilon$, near $z_2=0$,
the maximum of the Weyl transform of $\hPhi$ still defines the curve
$z_1=\epsilon\b z_2^2$, as we will soon see.

We can continue this procedure to higher orders.
At the $n^{th}$ order we will have an approximate
$\hLam_n$ that is correct up to (but not including) 
$\mathcal{O}(\epsilon^{n+2})$.
We can then calculate $\hPhi=e^{-i\hLam_n}\hP_1 e^{i\hLam_n}$
and define $\hat{\Xi}_n\equiv\com{\Delta\hPhi}{\hPhi}$ which will be of order
$\mathcal{O}(\epsilon^{n+3})$.
We can then try to correct $\hLam_n$ by a Hermitian operator
that will cancel $\hat{\Xi}_n$ up to the $(n+3)^{rd}$ order.
To find this we set $\hLam_{n+1}=\hLam_n+\delta\hLam$.
We then write the linearized equation for
$\delta\hPhi\equiv i\com{\hPhi_0}{\delta\hLam}+\mathcal{O}(\epsilon^{n+4})$.
It is:
\be\label{linpheq}
\com{\Delta\delta\hPhi}{\hPhi_0}+
\com{\Delta\hPhi_0}{\delta\hPhi} = -\Xi_n.
\ee
Here we can set:
$$
\hPhi_0 = \hP_1,\qquad \Delta\hPhi_0=\hadg{1}\hP_1\ha{1}-\hP_1.
$$
The equation (\ref{linpheq}) has solutions that are unique up to the zero
modes found above.
These are:
$$
\delta\hLam =
\sum_{n=0}^\infty C_n\hadg{1}\ha{2}^n
+\sum_{n=0}^\infty\overline{C}_n\ha{1}\hadgp{2}{n}
+\sum_{n=0}^\infty C'_n\hadg{1}\hadgp{2}{n}
+\sum_{n=0}^\infty\overline{C}'_n\ha{1}\ha{2}^{n}.
$$
We make sure that $\hLam$ does not contain these terms except
for the term $\epsilon\b\ha{2}^2$ that we began with.

At the next order we define $\hPhi=e^{-i\hLam_1}\hPhi_0 e^{i\hLam_1}$
and calculate $\Xi_2\equiv\com{\Delta\hPhi}{\hPhi}$.
We find:
$$
\Xi_2 =\frac{2}{3}\epsilon^4 (\b^3\barb\hadgp{1}{2}\hP_1\ha{2}^4
-\b\barb^3\hP_1\ha{1}^2\hadgp{2}{4}) + \mathcal{O}(\epsilon^5).
$$
We can correct this by augmenting $\hLam$ to:
\bear
\hLam_3 &\equiv&
\epsilon (\b\hadg{1}\ha{2}^2 +\barb\ha{1}\hadgp{2}{2})
-\frac{4}{3}\epsilon^3 (\b^2\barb\hadg{1}\hadg{2}\ha{2}^3
+\barb^2\b\ha{1}\hadgp{2}{3}\ha{2})
\nn\\
&&
-\frac{i}{3}\epsilon^4 
(\b^3\barb\hadgp{1}{2}\ha{2}^4-\barb^3\b\ha{1}^2\hadgp{2}{4}).
\nn
\eear

Continuing this procedure we find that, up to
$\mathcal{O}(\epsilon^8)$ terms, the following is a solution of the equations
of motion:
\bear
\hLam &=&
\epsilon\b\hadg{1}\ha{2}^2
+\epsilon\barb\ha{1}\hadgp{2}{2}
\nn\\ &&
-\left(
\frac{4}{3}\epsilon^3\b^2\barb
-\frac{56}{15}\epsilon^5\b^3\barb^2
+\frac{3872}{315}\epsilon^7\b^4\barb^3
\right)\hadg{1}\hadg{2}\ha{2}^3
\nn\\ &&
-\left(
\frac{4}{3}\epsilon^3\b\barb^2
-\frac{56}{15}\epsilon^5\b^2\barb^3
+\frac{3872}{315}\epsilon^7\b^3\barb^4
\right)\ha{1}\hadgp{2}{3}\ha{2}
\nn\\ &&
-\left(\frac{i}{3}\epsilon^4\b^3\barb
-\frac{139 i}{45}\epsilon^6\b^4\barb^2
\right)\hadgp{1}{2}\ha{2}^4
+\left(\frac{i}{3}\epsilon^4\b\barb^3
-\frac{139 i}{45}\epsilon^6\b^2\barb^4
\right)\ha{1}^2\hadgp{2}{4}
\nn\\ &&
+\left(\frac{191}{45}\epsilon^5\b^3\barb^2
-\frac{40121}{945}\epsilon^7\b^4\barb^3
\right)\hadg{1}\hadgp{2}{2}\ha{2}^4
\nn\\ &&
+\left(\frac{191}{45}\epsilon^5\b^2\barb^3
-\frac{40121}{945}\epsilon^7\b^3\barb^4
\right)\ha{1}\hadgp{2}{4}\ha{2}^2
\nn\\ &&
+\frac{142 i}{45} \epsilon^6\b^4\barb^2
\hadgp{1}{2}\hadg{2}\ha{2}^5
-\frac{142 i}{45}\epsilon^6\b^2\barb^4
\ha{1}^2\hadgp{2}{5}\ha{2}
\nn\\ &&
-\frac{176}{945}\epsilon^7\b^5\barb^2\hadgp{1}{3}\ha{2}^6
-\frac{176}{945}\epsilon^7\b^2\barb^5\ha{1}^3\hadgp{2}{6}
\nn\\ &&
-\frac{17162}{945}\epsilon^7\b^4\barb^3\hadg{1}\hadgp{2}{3}\ha{2}^5
-\frac{17162}{945}\epsilon^7\b^3\barb^4\ha{1}\hadgp{2}{5}\ha{2}^3
+\mathcal{O}(\epsilon^8).
\nn
\eear

Substituting this into $\hPhi$ and performing a Weyl transformation,
we can find the expression for the field $\Phi(z_1,z_2,\bz_1,\bz_2)$.
Since the expression is rather long, we will only present the leading 
terms below:
\bear
\Phi &=& 2e^{-|z_1|^2 +\psi},\nn\\
\psi &=&
\frac{\sqrt{2}}{2}i \epsilon (\b z_2^2\bz_1 -\barb z_1\bz_2^2)
+\epsilon^2 (-|\b|^2 +|\b|^2 z_1\bz_1
+2|\b|^2 z_1 z_2\bz_1\bz_2
-\frac{1}{2}|\b|^2 z_2^2\bz_2^2)
\nn\\ &&
+i\sqrt{2}\epsilon^3 (
-\frac{1}{6}\b|\b|^2z_2^2\bz_1
+\b|\b|^2 z_1 z_2^2\bz_1^2
-\b|\b|^2 z_2^3\bz_1\bz_2
\nn\\ &&
+\frac{1}{6}\barb|\b|^2 z_1\bz_2^2
-\barb|\b|^2 z_1^2\bz_1\bz_2^2
+\barb|\b|^2 z_1 z_2\bz_2^3)
\nn\\ &&
+\epsilon^4 (
\b^2|\b|^2 z_2^4\bz_1^2
+ \barb^2|\b|^2 z_1^2\bz_2^4
-\frac{17}{6}|\b|^4
+\frac{1}{3}|\b|^2 z_1\bz_1
\nn\\ &&
+|\b|^2 |z_1|^4
+6|\b|^2 |z_2|^2
-\frac{16}{3}|\b|^4 |z_1|^2 |z_2|^2
+4 |\b|^4 |z_1|^4 |z_2|^2
\nn\\ &&
-\frac{1}{6}|\b|^4 |z_2|^4
-8|\b|^4 |z_1|^2 |z_2|^4
+|\b|^4 |\bz_2|^6
)
+\mathcal{O}(\epsilon^5).
\label{SolUpTo7}
\eear

We can now look for the maximum of $\Phi$. This will approximately
outline the curve that a macroscopic observer would see as a 2-brane.
The minimum of the exponent is at:
\bear
z_1 &=& \frac{i}{\sqrt{2}}\epsilon\b z_2^2 (
1+\frac{2}{3}\epsilon^2|\b|^2+\frac{188}{15}\epsilon^4|\b|^4
+\frac{8956}{315}\epsilon^6|\b|^6
-96\epsilon^6|\b|^6 |z_2|^2) + \mathcal{O}(\epsilon^8).
\label{SolZ}
\eear
We see that up to order $\mathcal{O}(\epsilon^6)$, all the corrections
can be interpreted as a renormalization of $\b$. At large scale
there are no corrections to the parabolic shape of the graph of the brane.
In particular, the curve is still analytic.
The first deviation from analyticity occurs at order 
$\mathcal{O}(\epsilon^7)$ because
of the appearance of the $z_2^2|z_2|^2$ term.
To this order $\varphi$ is no longer harmonic and instead satisfies:
$$
z_1 =\varphi(z_2,\bz_2),\qquad
\pxz\bpx\varphi = 
-18(\pxz\varphi)^2(\pxz^2\varphi)^2(\bpx^2\overline{\varphi})^3.
$$

\subsection{Region of Validity of the Approximation}
We have started to construct, order by order, a solution that
looks macroscopically near the origin like the curve
$z_1=\epsilon\b z_2^2$.
By ``macroscopically'' we mean that distances are larger than the noncommutativity
scale. We have set the noncommutativity scale to $1$ here, so we require
that the solution be valid not only for $z_1,z_2\sim 0$ but also for
$|z_1|,|z_2|\gg 1$!
On the other hand, we wish to assume that the curvature of the curve is small
at the origin and, as far as the geometry of the curve goes, we
are in the vicinity of the origin.
Quantitatively, this requires that $|\epsilon\b z_2|\ll 1$.
Looking at the solution, (\ref{SolUpTo7}), we see that the
order of magnitude of the $\mathcal{O}(\epsilon^{2n})$ in $\hLam$ is smaller by
a factor of $\epsilon\b z_1$ from the $\mathcal{O}(\epsilon^{2n-1})$ terms
and the $\mathcal{O}(\epsilon^{2n+1})$ terms are smaller by a factor
of $\epsilon^2|\b|^2|z_2|^2$ from the $\mathcal{O}(\epsilon^{2n-1})$ terms.
So, the approximation is within the required region of validity.

Note, however, that in the region of validity of the calculation,
{\it ie}, $\epsilon|\b z_2|\ll 1$, the correction to $z_1$ in
(\ref{SolZ}) is smaller than $1$,
and thus is actually microscopic.

\section{Intersecting D2-Branes}
\label{twobr}

\subsection{Construction of the Intersecting Soliton}

In the previous section, we constructed a D2-brane at $z_1=0$ as
$\hPhi_1=\lambda \hP_1$ and a D2-brane at $z_2=0$ as
$\hPhi_2=\lambda \hP_2.$
We now wish to find a soliton $\hPhi = \lambda \hP$
which asymptotically looks like $\hPhi_1 + \hPhi_2$.
This is straightforward. We define
$$
\hP_\dl = \hP_1 + \hP_2 -\dl \hP_1 \hP_2,
\qquad \hPhi_\dl = \lambda \hP_\dl
$$
This will be a projection operator for $\dl  =1$ or $\dl=2$.
To distinguish between the two solutions, we need to calculate
their kinetic energy, (\ref{kehilb}).
While each solution has an infinite kinetic energy
because of its infinite extent, the difference
is finite and easy to calculate:
$$
K(\hPhi_{\dl=2}) -K(\hPhi_{\dl=1}) = \frac{4 \lambda^2}{\th}.
$$
Thus, $\dl=1$ corresponds to the solution with the lower
kinetic energy.  We propose that this solution corresponds to
two intersecting branes.  The $\dl=2$ solution is similar,
but it has a `hole' attached at the intersection:
$$
\hP_{\dl=2} = \hP_{\dl=1} - \hP_1 \hP_2.
$$
In a sense, it is as if a 0-brane (represented by $\hP_1 \hP_2$)
had been removed.  This
solution will turn out to be unstable to small unitary perturbations.

\subsection{Fluctuations}
We now wish to repeat the calculation of the
effective Hamiltonian for small fluctuations
of the two intersecting branes.
Consider the fluctuation given by $\hU^{\dagger} \hP_{\dl} \hU$,
where $\hU$ is again a unitary operator on $\Hil = \Hil_1 \otimes \Hil_2.$
As before, let
\be\label{Ulam}
\hU = e^{i\epsilon\hLam} = 1 + i\epsilon \hLam -
\frac{1}{2} \epsilon^2 \hLam^2 + \mathcal{O}(\epsilon^3)
\ee
with $\epsilon$ real and small and $\hLam$ hermitian. One can
calculate the kinetic energy for this soliton to second order in
$\epsilon$.  This is most conveniently done from equation (\ref{kehilb}).

In the $\dl = 1$ case, we define
\bear
\hLam \ket{0,j} &=& \sum_{I,i} b^j_{Ii} \ket{I,i},
\quad I,i,j\ge 1,
\nn\\
\hLam \ket{J,0} &=& \sum_{I,i} c^J_{Ii} \ket{I,i},
\quad I,i,J\ge 1
\nn\\
\hLam \ket{0,0} &=& \sum_{I,i} d_{Ii} \ket{I,i}.
\quad I,i\ge 1
\eear

After consolidation of terms, (\ref{kehilb}) becomes:
\bear
\frac{\theta} {2 \lambda^2}  K_{\dl=1}&=&
\frac{\theta } {2 \lambda^2} K(\hPhi_{\dl=1})+
2\epsilon^2 \left [
\sum_{J \ge 2, j,k \ge 1} J |b^k_{Jj}|^2 +
\sum_{j \ge 2, J,K \ge 1} j |c^K_{Jj}|^2 \right.
\nn\\ &+&
\sum_{J,j,k \ge 1}
\left|\sqrt{k+1} b^k_{Jj} - \sqrt{j+1} b^{k+1}_{J,{j+1}}\right|^2
+
\sum_{J,j,K \ge 1}
\left|\sqrt{K+1} c^k_{Jj} - \sqrt{J+1} c^{K+1}_{{J+1},j}\right|^2
\nn\\ &+&
\sum_{J \ge 2, j \ge 1} J |d_{Jj}|^2 +
\sum_{j \ge 2, J \ge 1} j |d_{Jj}|^2
\nn\\ &+&
\left .
\sum_{J,j \ge 1} \left|d_{Jj} - \sqrt{j+1} b^1_{J,{j+1}}\right|^2 +
\sum_{J,j \ge 1} \left|d_{Jj} - \sqrt{J+1} c^1_{{J+1},j}\right|^2
\right ]
\eear
where $K(\hPhi_{\dl=1})$ is the (infinite) energy
of an undistorted soliton discussed in previous subsection.

Using the same procedure as before, we obtain the following
zero modes:
$$
b^m_{1,n} = \left\{\begin{array}{ll}
0 & {\mbox{for $m+1 < n$}} \\
\frac {d_{11}}{\sqrt{n}} & {\mbox{for $m+1 = n$}} \\
d_{10} & {\mbox{for $m = n$}} \\
\sqrt{\frac{m!}{n!}} p_{m-n} &
     {\mbox{for $m > n$}} \\
\end{array}\right.
$$
and
$$
c^M_{N,1} = \left\{\begin{array}{ll}
0 & {\mbox{for $M+1 < N$}} \\
\frac {d_{11}}{\sqrt{N}} & {\mbox{for $M+1 = N$}} \\
d_{01} & {\mbox{for $M = N$}} \\
\sqrt{\frac{M!}{N!}} q_{M-N} &
     {\mbox{for $M > N$}} \\
\end{array}\right.
$$
with $d_{Jj}$ for all $J,j \ge 2$ equal to zero.
The $p$'s and $q$'s are arbitrary constants.  These can be used to define
two entire holomorphic functions
$f_1(\zeta) = \sum_m p_m \zeta^m$ and
$f_2(\zeta) = \sum_M q_M \zeta^M.$ These zero modes, just as for a
single brane,
correspond to deformations of the two branes:
$z_1' =  \epsilon f_1(z_2)$ and $z_2' = \epsilon f_2(z_1).$
As in the case of a single brane, the terms in the kinetic energy can be
rearranged to make apparent the antiholomorphic deformations.

A new phenomenon is the mode corresponding to a
non-zero $d_{11}$ together with
$b^k_{1,k+1}=d_{11}(k+1)^{-1/2}$ and
$c^K_{K+1,1}=d_{11}(K+1)^{-1/2}$
so that the terms in kinetic energy that are differences
vanish.  This mode might be thought of as
$$
\Lambda \sim  \frac{\alpha}{z_1 z_2} +
\frac{\bar \alpha}{z^{\dagger}_1 z^{\dagger}_2}.
$$
This is a complex mode (two real modes) corresponding to the
extra degrees of freedom living on the intersection of the two
branes.

We now consider the case of $\dl=2$. Here, (\ref{kehilb}) reduces to
\bear
\frac{\theta} {2 \lambda^2}  K_{\dl=1}&=&
\frac{\theta} {2 \lambda^2}  K(\hPhi_{\dl=1})+
2\epsilon^2 \Bigg [
\sum_{J \ge 2, j,k \ge 1} J |b^k_{Jj}|^2 +
\sum_{j \ge 2, J,K \ge 1} j |c^K_{Jj}|^2
\nn\\ &+&
\sum_{J,j,k \ge 1}
\left|\sqrt{k+1} b^k_{Jj} - \sqrt{j+1} b^{k+1}_{J,{j+1}}\right|^2
+
\sum_{J,j,K \ge 1}
\left|\sqrt{K+1} c^k_{Jj} - \sqrt{J+1} c^{K+1}_{{J+1},j}\right|^2
\nn\\ &+&
\sum_{J,k \ge 2} (J+k-1) |b^1_{Jk}|^2 +
\sum_{J \ge 2} (J-1) |b^1_{J+1,1}|^2 +
\sum_{k \ge 2} (k-1) |b^1_{1,k}|^2
\nn\\ &+&
\sum_{j,K \ge 2} (j+K-1) |c^1_{Kj}|^2 +
\sum_{j \ge 2} (j-1) |c^1_{1,j+1}|^2 +
\sum_{K \ge 2} (K-1) |c^1_{K,1}|^2
\nn\\ &+&
\sum_{k \ge 2} k |b^{k+1}_{00}|^2 +
\sum_{K \ge 2} K |c^{K+1}_{00}|^2 +
\sum_{K \ge 2} |b^{1}_{K1} + \bar c^K_{00}|^2 +
\sum_{k \ge 2} |c^{1}_{1k} + \bar b^k_{00}|^2
\nn\\ &-&
\left|b^{1}_{11} + \bar c^1_{00}\right|^2 -
\left|c^{1}_{11} + \bar b^1_{00}\right|^2
\Bigg].
\eear
The zero modes, which we will not write out explicitly,
include our familiar entire holomorphic and anti-holomorphic deformations
of the branes. More importantly, we now have unstable modes
given by $b^1_{11} + \bar c^1_{00} \ne 0$ and
$c^1_{11} + \bar b^1_{00} \ne 0$ together with extra elements
so that the terms that are differences are zero.
These two modes correspond to moving the aforementioned `hole'
away from the intersection along either of the two branes.
We also note that the above effective Hamiltonian has an
additional zero mode given by
$\Lambda = \a(\hadg{1})^2 + \bar{\a} (\ha{1})^2$
(and similarly for $\ha{2}$), which corresponds to distorting the
shape of the hole from the gaussian ground state of a harmonic oscillator
into a squeezed state.

\section{Extensions and Discussion}
\label{disc}

\subsection{Multiple Branes}
Our construction for two intersecting D2-branes can easily
be extended to a larger number of branes.

For example, let
$\hP^K_1$ be a projection operator corresponding to a stack of
K branes at $z_1=0$ and $\hP^L_2$ be a projection operator
corresponding to a stack of L branes at $z_2=0$.
This means that $\hP^K_1$ can be written as
a sum of $K$ projection operators
$$
\hP^K_1 = \sum_{i=1}^{K}\hp^i_1
$$
with $\hp^i_1 \hp^j_1 = \delta^{ij} \hp^i_1$, each $\hp^i_1$ being
a projection operator for a single brane.
Similarly,
$$
\hP^L_2 = \sum_{i=1}^{L}\hp^i_2.
$$
Now, any operator of the form
$$
\hP^K_1 + \hP^L_2 - \sum_{i=1}^{K}\sum_{j=1}^{L}
\dl_{ij} \hp^i_1 \hp^j_2
$$
for $\dl_{ij}=1,2$ corresponds to an intersection of these two stacks.

As another example, let us take $\MR{6}$, i.e. three complex dimensions.
Let $\hP_{12}$ correspond to a codimension-2 brane at $z_3=0$,
$\hP_{23}$ correspond to a codimension-2 brane at $z_1=0$ and
$\hP_{31}$ correspond to a codimension-2 brane at $z_2=0$.
Then it can be checked that
$$
\hP_{12} + \hP_{23} + \hP_{31}
- \dl_{12} \hP_{23} \hP_{31}
- \dl_{23} \hP_{12} \hP_{31}
- \dl_{31} \hP_{12} \hP_{23}
+ (\dl_{12}+\dl_{23}+\dl_{31} - \dl - 1)
\hP_{12} \hP_{23} \hP_{31}
$$
is a projection operator corresponding to the intersection
of all three branes at a point, provided we set
$\dl_{12},\dl_{23},\dl_{31},\dl\in\{1,2\}$.
It is straightforward, if a bit tedious, to extend this to any number
of branes.

\subsection{Discussion}
In this paper we have found a solution that describes two intersecting
2-brane solitons in a field theory on a noncommutative $\MR{4}$ in
the large noncommutativity limit. We studied the zero modes of the
solution. We found a zero mode that is reminiscent of the zero mode
of two intersecting D2-branes that corresponds to a  deformation
into an irreducible curve.
It would be interesting to examine whether this zero mode receives
a potential at higher orders or whether it is an exact flat direction.
Because we have seen that simple holomorphic deformations are no
longer flat at sufficiently high orders, the latter possibility seems unlikely.
Here ``higher-order'' could
have several meanings. First there is the expansion of the classical
action, still in the large noncommutativity limit. This expansion
parameter is the $\epsilon$ in equation (\ref{Ulam}).
On top of that, there are the expansions in the noncommutativity parameter
and the quantum fluctuations. In these notes we have not attempted to
include either of those.

It has recently been shown \cite{GroNek,AGMS,HKLii} that, in the
situation of a noncommutative $U(\infty)$ gauge theory, one can cancel
the kinetic term in the action through a suitable configuration of the
gauge fields. The soliton configurations discussed in this paper are
easily realizable in the schemes of the referenced papers. It might
be interesting to compute the actions for perturbations of the fields
as in \cite{GroNek,AGMS}. However, because one can always find a gauge
field to cancel the kinetic term of a given soliton and because the
projection operators here are halving projections\footnote{Both $\hP$
and $1-\hP$ have infinite rank. All such projection operators are
unitarily equivalent.}, it should be
possible to continuously interpolate through conjugation with unitary
operators between these soliton
configurations and other configurations that are described
by halving projections. This includes, in the case of four
noncommutative directions, any number of branes in any given direction.
It is not immediately clear to us what this means.


\section*{Acknowledgments}
We are grateful to Keshav Dasgupta, Govindan Rajesh and Leonardo
Rastelli for helpful discussions.
We also wish to thank Natalia Saulina for participating in early
stages of this project.
This research is supported by NSF grant number PHY-9802498.
The research of JLK is in part supported by the Natural Sciences and
Engineering Research Council of Canada.

\def\np#1#2#3{{\it Nucl.\ Phys.} {\bf B#1} (#2) #3}
\def\pl#1#2#3{{\it Phys.\ Lett.} {\bf B#1} (#2) #3}
\def\physrev#1#2#3{{\it Phys.\ Rev.\ Lett.} {\bf #1} (#2) #3}
\def\prd#1#2#3{{\it Phys.\ Rev.} {\bf D#1} (#2) #3}
\def\ap#1#2#3{{\it Ann.\ Phys.} {\bf #1} (#2) #3}
\def\ppt#1#2#3{{\it Phys.\ Rep.} {\bf #1} (#2) #3}
\def\rmp#1#2#3{{\it Rev.\ Mod.\ Phys.} {\bf #1} (#2) #3}
\def\cmp#1#2#3{{\it Comm.\ Math.\ Phys.} {\bf #1} (#2) #3}
\def\mpla#1#2#3{{\it Mod.\ Phys.\ Lett.} {\bf #1} (#2) #3}
\def\jhep#1#2#3{{\it JHEP.} {\bf #1} (#2) #3}
\def\atmp#1#2#3{{\it Adv.\ Theor.\ Math.\ Phys.} {\bf #1} (#2) #3}
\def\jgp#1#2#3{{\it J.\ Geom.\ Phys.} {\bf #1} (#2) #3}
\def\cqg#1#2#3{{\it Class.\ Quant.\ Grav.} {\bf #1} (#2) #3}
\def\hepth#1{{\tt hep-th/{#1}}}


\end{document}